\documentclass[twocolumn,showpacs,showkeys,superscriptaddress]{revtex4}
\usepackage{graphicx}
\usepackage{bm}
\usepackage{color}
\usepackage{amsmath}
\usepackage{amsfonts}
\usepackage{amssymb}
\usepackage{natbib}
\usepackage{latexsym}

\begin{document}

\title{Parametric amplification of electromagnetic plasma waves in resonance with a  dispersive  background gravitational wave }
\author{Swadesh M. Mahajan}
\email{mahajan@mail.utexas.edu}
\affiliation{Institute for Fusion Studies, The University of Texas at Austin, Austin, Texas 78712, USA.}
\author{Felipe A. Asenjo}
\email{felipe.asenjo@uai.cl}
\affiliation{Facultad de Ingenier\'{\i}a y Ciencias, Universidad Adolfo Ib\'a\~nez, Santiago 7941169, Chile.}


\begin{abstract}
It is shown that a sub-luminal electromagnetic plasma wave, propagating in phase with a background sub-luminal gravitational wave in a dispersive medium, can undergo parametric amplification. For this  phenomena to occur, the dispersive characteristics of the two waves must properly match. The response frequencies of the two waves (medium dependent) must lie within a definite and restrictive range. The combined dynamics is represented  by a Whitaker-Hill equation, the quintessential model for parametric instabilities. The exponential growth of the electromagnetic wave is displayed at the  resonance; the plasma wave grows at the expense of the background gravitational wave.  Different physical scenarios, where the phenomenon can be possible, are discussed.

\end{abstract}

\pacs{}
\keywords{ }

\maketitle

\section{Introduction}

Resonant interactions between two distinct waves, controlled by the same wave operators, are expected to be especially strong. In a series of recent papers, it was shown that hyperbolic waves, in particular the dispersive electromagnetic and the gravitational waves,  could very efficiently transfer energy to relativistic electrons described by a Klein-Gordon wave \cite{mahajan1, mah2, asenjo3}. These semiclassical calculations (Klein-Gordon equation representing a quantum relativistic spin less particle \cite{footnote}) demonstrated that  both electromagnetic and gravitational waves, through wave-wave interaction, could resonantly accelerate relativistic electrons to high energies. The resonance occurs when the energy propagation speed of the Klein-Gordon wave and the classical (electromagnetic or gravitational)  wave are equal. This is possible only when the classical wave is propagating in a dispersive medium and its group velocity is close to but less than unity. 

This work is an enquiry into what happens when such a mildly sub-luminal electromagnetic wave propagates in a plasma medium embedded in curved spacetime, specifically, when the space-time curvature is due to a similarly sub-luminal background dispersive gravitational wave. We show that the resonant interaction rises to an altogether different level of interest; the phenomenon of parametric amplification of the electromagnetic wave is observed. That is, when the dispersive properties of the the two waves (in the absence of the other) bear specific quantitative relations to one another, even a very small amplitude gravitational wave can drive an electromagnetic plasma wave to large amplitudes. An important and necessary requirement  for this phenomenon to occur is that the background gravitational wave must be dispersive (possible in a background massive medium) \cite{forsberg,brodin,mendonca,bamba,asseo,chesters,gayer,polnarev,ingram,servin,mendonca2,silker}. Only then the gravitational wave becomes subluminal, and thus,  couples to the subluminal  electromagnetic field  in a resonant manner in a plasma.

In a general context, parametric instabilities occur in a variety of physical systems with periodic potentials.  Perhaps the most familiar manifestation is the existence of the the band structure (Brillouin zones) in metals, derived by solving the Schr\"odinger equation for an electron moving in the periodic ionic background \cite{Pettifor}. Parametric resonance is an ubiquitous effect  in every branch of physics, such as, for example, in plasmas \cite{Yamanaka,Mamun,Smirnov}, lasers \cite{Lin,Krokhin,Huang,Drouet,Barr,Arefiev,Kuzanyan}, cosmology \cite{AZZ,Paulo}, astrophysics and particle physics \cite{Figueroa,Akhmedov,Zlatev}, etc.
Though the wave-wave interaction and energy exchange processes permeate all physics, the phenomenon reported here is rarer. It is possible when the two waves are resonant and, simultaneously, satisfy certain rather ``rigid" relationships in their dispersion characteristics, especially when the amplitude of the driving gravitational wave is low. These limiting features of the parametric process will be later discussed when the coupled equations are analyzed.

Overall there are different studies on linear and nonlinear resonant energy exchange processes 
 in which  plasma waves in curved spacetime (Alfven waves, cyclotron waves, etc.) grow at the cost of gravitational waves \cite{forsberg,forsberg2,brodinPRL,servin62}, or particles can be accelerated by gravitational waves \cite{Loeb, Papadopoulos,Papadopoulos2,Papadopoulos3,Papadopoulos4}. A direct resonant amplification of plasma waves due to a light-like (non-dispersive) gravitational wave was studied in Ref.~\cite{brodinPRL}. However, to the best of our knowledge, no similar process for a subluminal gravitational wave has been proposed. In this case, as we will see below, the parametric resonance occurs when the dispersive properties of the systems are
very specific.

Although the essential physics behind parametric amplification will be accessible in a rather simple equation, we will  begin by deriving in Sec. II  the exact equations obeyed by transverse electromagnetic plasma waves propagating in a plasma immersed in a curved spacetime background. In Sec. III, we explore the solutions for those electromagnetic waves when the curvature is due to a gravitational wave background. We will, then, go on to investigate the rather simplified equation (pertaining to small amplitude gravitational wave). We will, first, extract the precise conditions when parametric amplification is possible, and then, to calculate the characteristics of the enhanced electromagnetic plasma wave. In Sec. IV, we will discuss the implications of this work.

\section{Electromagnetic plasmas waves in general curved spacetime background}

 Our model system consists of a fluid plasma, immersed in a gravitational field, that  has one dynamic charged component (electrons) moving in a neutralizing background (provided by ions, for example). Such
an ideal general relativistic one-component plasma fluid (with charge $q$, mass $m$ and density $n$) can be described in a unified form as \cite{mahajan, asenjo2}
\begin{equation}\label{plasanifs}
q U_\nu M^{\mu\nu}=T \nabla^\mu\sigma\, .
\end{equation}
where $\nabla_\mu$ is a covariant derivative for a metric $g_{\mu\nu}$ with signature $(-,+,+,+)$, and with $\mu,\nu=0,1,2,3$. Here, 
 $U^\mu$ is the plasma fluid four-velocity, $T$ is the plasma temperature, $\sigma$ is its entropy, and 
\begin{equation}
M^{\mu\nu}=F^{\mu\nu}+\frac{m}{q} S^{\mu\nu}\, ,
\end{equation}
is the  magnetofluid tensor  unifying the electromagnetic field, described by the tensor $F^{\mu\nu}=\nabla^\mu A^\nu-\nabla^\nu A^\mu$ (with the electromagnetic four-potential
$A^\mu$), and the fluid vorticity antisymmetric tensor
$S^{\mu\nu}=\nabla^\mu \left(f U^\nu\right)-\nabla^\nu \left( f U^\mu\right)$, with $f=h/mn$ , where $h$ is the plasma enthalpy density.  Here, $f$ contains the thermal-inertial effects of the plasma, and it can be assumed as $f=K_3[(m c^2)/(k_B T)]/K_2[(m c^2)/(k_B T)]$,
where $K_2$
and $K_3$ are the modified Bessel functions of the second
kind of orders 2 and 3, $c$ is the speed of light, and $k_B$ is the Boltzmann constant \cite{mahajan, asenjo}. The above description is valid for an isentropic plasma,  as  $U_\nu \nabla^\nu\sigma=0$. This plasma dynamics will provide the four current needed to close the system through Maxwell equations
\begin{equation}\label{maxwell}
\nabla_\nu F^{\mu\nu}=q n U^\mu\, .
\end{equation}

For a homentropic plasma fluid $\nabla^\mu\sigma=\partial^\mu\sigma=0$, and electromagnetic plasma waves are straightforward solution for a plasma in any curved spacetime. In fact, the dynamics is reduced to $M^{\mu\nu}=0$ that leads to a simple relationship between the transverse components ($\mu=1,2$) of the current and the vector potential,  
\begin{equation}\label {Current-Pot}
A^\mu+\frac{m f}{q} U^\mu=0\, ,
\end{equation}
Substituting \eqref{Current-Pot} into \eqref{maxwell}, we arrive at the wave equation 
\begin{equation}
\nabla_\nu F^{\mu\nu}+\Omega_p^2 A^\mu=0\, .
\label{plasmawaveequation}
\end{equation}
 where $\Omega_p=\omega_p/\sqrt{f}$ is a constant, with the plasma frequency $\omega_p=\sqrt{n q^2/m}$. This equation can describe electromagnetic plasma waves in any curved spacetime.

In curved spacetime, Gravity will enter Eq.~\eqref{plasmawaveequation} through the tensor derivatives. The propagation characteristics of the electromagnetic waves, then,  will depend on the background curved spacetime in addition to the dielectric properties of the medium that enter through the plasma frequency (and make the propagation sub luminous). In the following section, we will show that, under appropriate conditions, the background curved spacetime provided by a dispersive gravitational wave is able to trigger an instability driving the electromagnetic wave to high amplitudes.

\section{Parametric resonance due to dispersive gravitational wave background}

A gravitational wave  propagating in a massive medium, is dispersive  and subluminal;
the medium is endowed with a refractive index \cite{forsberg,brodin,mendonca,bamba,asseo,chesters,gayer,polnarev,ingram,servin,mendonca2,silker}. 
An alternative interpretation is that the graviton acquires an effective mass in a massive medium just like the photon does in a plasma. 

Let us model a dispersive gravitational wave as a a perturbation on the flat spacetime. Without loss of generality, let us assume that it is propagating in the $z$-direction. The spacetime  interval $ds^2=g_{\mu\nu}dx^\mu dx^\mu$ ($\mu, \nu=0,1,2,3$), described by the metric $g_{\mu\nu}=\eta_{\mu\nu}+h_{\mu\nu}$ is split into $\eta_{\mu\nu}=(-1,1,1,1)$, the flat spacetime metric, and $h_{\mu\nu}$  the  perturbation ($h_{\mu\nu}\ll \eta_{\mu\nu}$).  Let us further restrict to a simple gravitational wave with two nonzero components $h_{11}=-h_{22}=h(\chi)$ that are functions only of the wave phase, $\chi=\omega t- k z$, where $\omega$ and $k$ being, respectively the frequency and wavenumber \cite{forsberg,brodin,mendonca,bamba,asseo,chesters,gayer,polnarev,ingram,servin,mendonca2,silker}.  

The simplest example of a dispersive gravitational wave has a dispersion relation of the form $\omega^2-k^2\equiv\omega_G^2\neq 0$ (in close analogy with the electromagnetic wave in a plasma) where $\omega_G$ (determined by the properties of the medium) forces its group velocity ($d\omega/dk$) to fall below unity. In general, it is expected that the response frequency be quite small compared to the frequency of the wave,
$\omega_G\ll\omega,k$; the gravitational wave will travel with group velocities near (but below) the speed of light.


We are now ready to go back to seek more explicit solutions of Eq.~\eqref{plasmawaveequation}. Spelling out the tensor derivatives (notice that the amplitude of the electromagnetic wave has to be small enough that it does not affect gravity), we have
\begin{equation}\label{electroplasmaec}
\frac{1}{\sqrt{-g}}\partial_\nu\left[\sqrt{-g}g^{\mu\alpha}g^{\nu\beta}\left(\partial_\alpha A_\beta-\partial_\beta A_\alpha  \right)\right]+\Omega_p^2 g^{\mu\alpha}A_\alpha=0\, .
\end{equation}
Fixing $z$ to be the direction of propagation, we let the transverse components of the electromagnetic wave potential to be $A_1=A_+$ and $A_2=A_-$. Using the metric for the gravitational wave (specified in previous paragraph), these potentials obey  
\begin{equation}
\frac{\partial}{\partial t}\left(f_\pm \frac{\partial A_\pm}{\partial t} \right)-\frac{\partial}{\partial z}\left(f_\pm \frac{\partial A_\pm}{\partial z} \right)+\Omega_p^2 f_\pm A_\pm=0\, ,
\label{plasmawaveequation2}
\end{equation}
where $f_+=\sqrt{-g}g^{11}\approx 1-h$,  and  $f_+=\sqrt{-g}g^{22}\approx 1+h$. 

Since we are out to explore the interaction of resonant electromagnetic plasma waves and gravitational waves, we demand that we seek solutions in which the electromagnetic potentials are functions of exactly the same phase as  gravitational waves, i.e,  $A_\pm=A_\pm(\chi)$. In this form, the phase velocity of the electromagnetic plasma waves coincide with the phase velocity of the gravitational wave background.
The wave equation \eqref{plasmawaveequation2}, then, becomes an ordinary differential equation 
 \begin{equation}
\frac{d^2A_\pm}{d\chi^2}\mp \frac{d h}{d\chi}\frac{d A_\pm}{d\chi}+\left(\frac{\Omega_p}{\omega_G}\right)^2A_\pm=0\, ,
\label{ecGenAhi}
\end{equation}
representing a driven homogeneous oscillator; the driver being the gravitational field $h$. 

We can consider a gravitational wave described by a plane wave form $h=h_0\cos\chi$, with the respective phase $\chi$ and amplitude $h_0\ll 1$. In this case, Eq.~\eqref{ecGenAhi} can be write it more explicitly as
\begin{equation}
\frac{d^2A_\pm}{d\chi^2}\pm h_0 \sin\chi \frac{d A_\pm}{d\chi}+\left(\frac{\Omega_p}{\omega_G}\right)^2A_\pm=0\, .
\label{ecGenAh}
\end{equation}
The oscillator is clearly subject to a periodic potential; this has immensely interesting consequences.
Noting that $A_-(\chi)=A_+(\chi-\pi)$, it is enough to solve for only one polarization. For the $A_-$ polarization, a change of variable 
\begin{equation}
{\cal A}_-=\exp\left(\frac{h_0}{2}\cos\chi\right) A_-\,  ,
\end{equation}
converts Eq.~\eqref{ecGenAh} into the more standard form of
 a Whittaker-Hill equation, 
\begin{equation}
\frac{d^2{\cal A}_-}{d\chi^2}+\left[ \left(\frac{\Omega_p}{\omega_G}\right)^2+ \frac{h_0}{2}\cos\chi-\frac{h_0^2}{4}\sin^2\chi\right]{\cal A}_-=0\, ,
\label{ecGenAhWH}
\end{equation}
A similar equation can be found for $+$ polarization.
There is vast literature on the exact solutions of equations like Eq.~\eqref{ecGenAhWH} with periodic coefficients.  However our goal here is  to explore the interesting physics and delineate the precise conditions where the parametric resonance is triggered. The key is to recognize  that the Whittaker-Hill equation has different instabilities zones for its two-dimensional  parameters  space $(\Omega_p/\omega_G, h_0)$  \cite{arscott}.

For small enough $|h_0|$, however, most zones of instability shrink, and mostly  the only regime where we can observe  parametric amplification is the first one. In this limit, neglecting high order terms $\mathcal{O}(h^2_0)$,  Eq.~\eqref{ecGenAhWH} reduces to a Mathieu  equation, where the first and likely the only relevant unstable region for the dynamics  is in the vicinity 
\begin{equation}
 \left(\frac{\Omega_p}{\omega_G}\right)^2=\frac{1}{4} \, .
\label{condition14}
\end{equation}
and has a width of order $h_0$ (see for instance, Ref.~\cite{bender}). In this part of the parameter space,
the electromagnetic plasma wave  displays an exponentially growing amplitude with the form
\begin{equation}
|{A}_-|\propto \exp\left(\frac{h_0}{4}\chi\right)\, .
\label{eqexornt}
\end{equation}
Since the $A_+$ polarization is just phase shifted, it will have the same exponential growth.

This parametric amplification is a  manifestation of a combination of two effects: the resonance phase between the electromagnetic plasma wave and the gravitational wave, and their dispersion characteristics having (as remarked earlier) a definitive relationship. Consequently, the electromagnetic plasma wave (of both polarizations) draws  energy, very efficiently, from the background gravitational field. In this way, the phenomenon described here has a pure general-relativistic origin.

\begin{figure}[ht]
\includegraphics[width =3.4in]{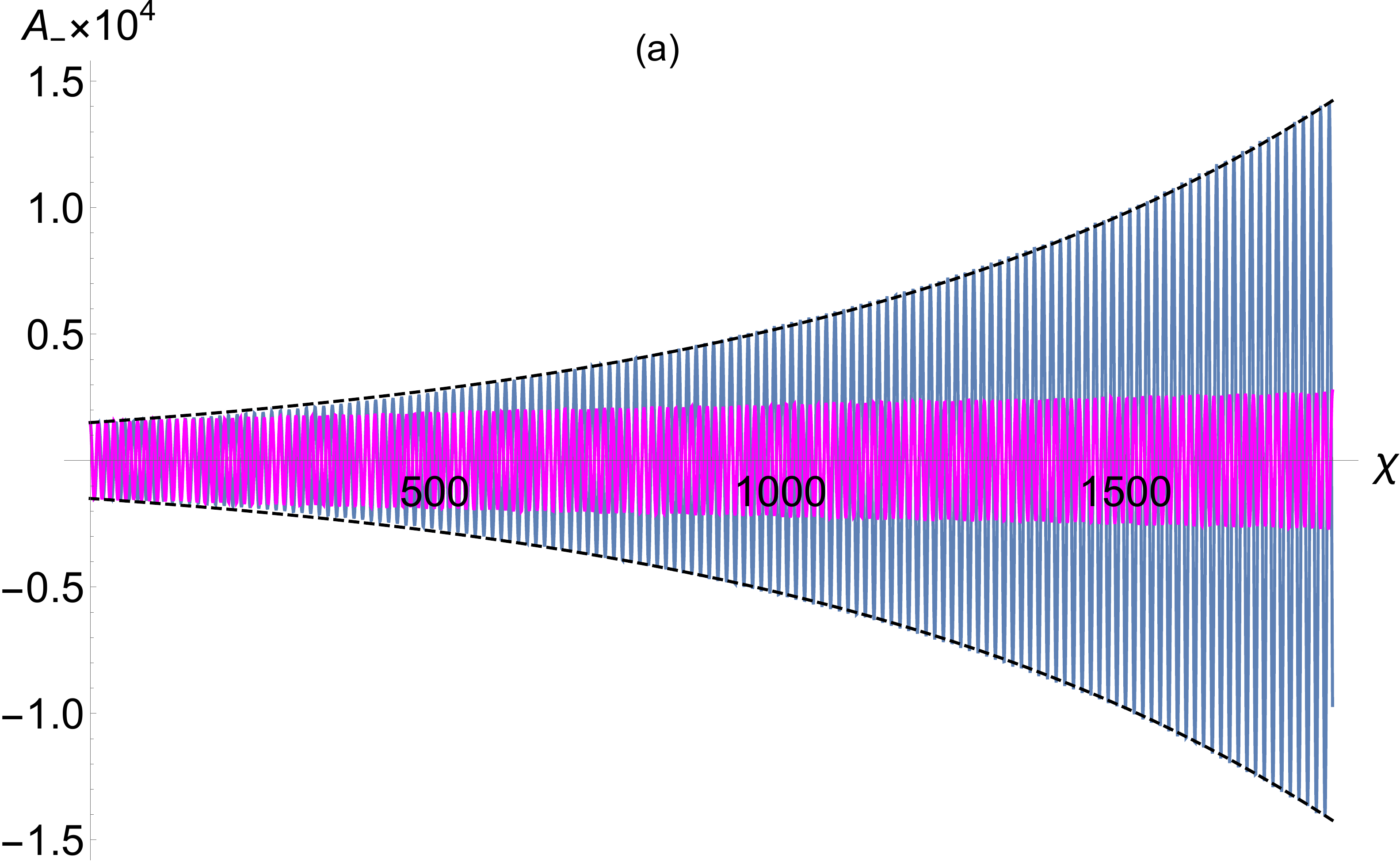}
\includegraphics[width =3.4in]{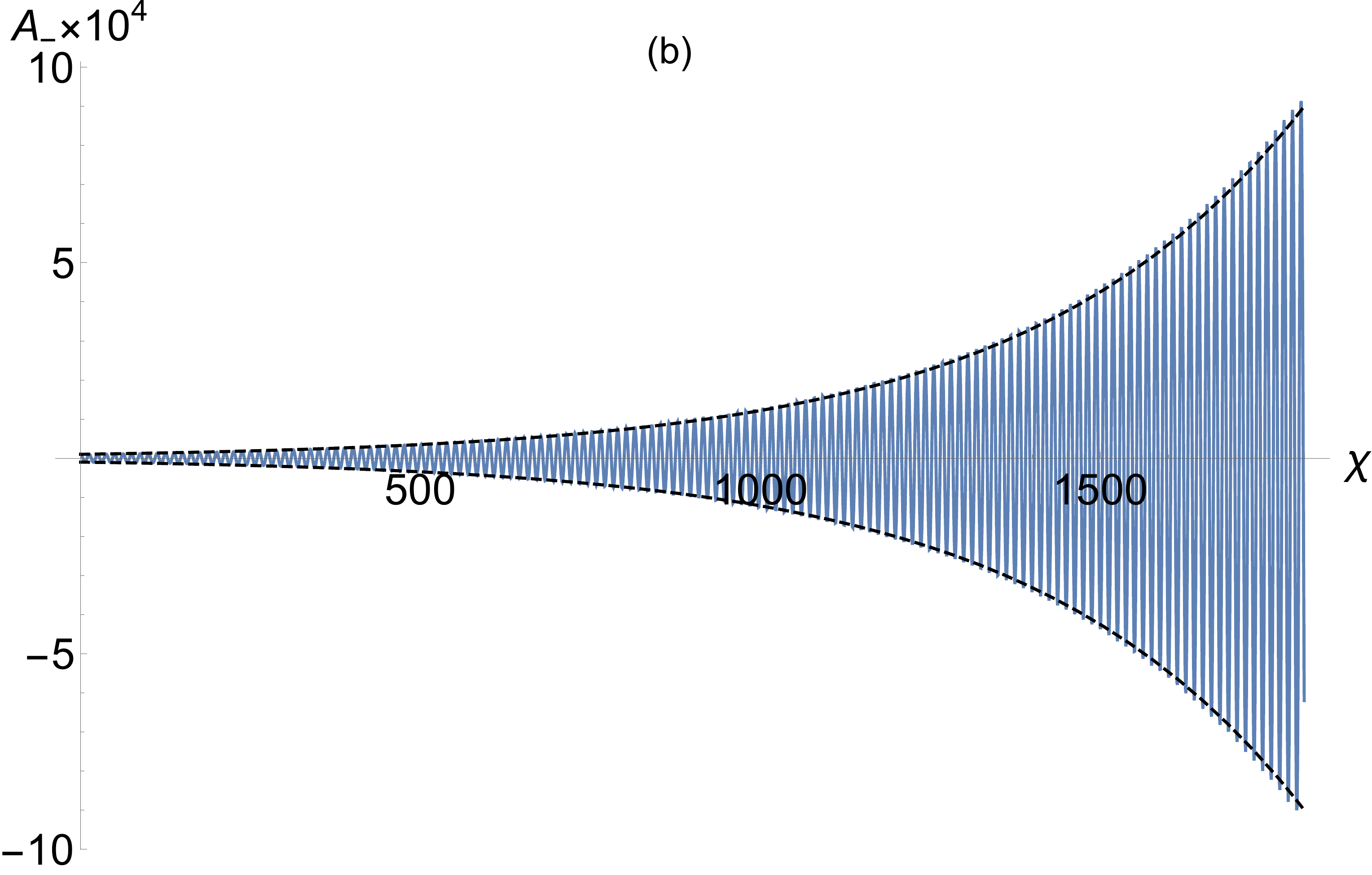}
\includegraphics[width =3.4in]{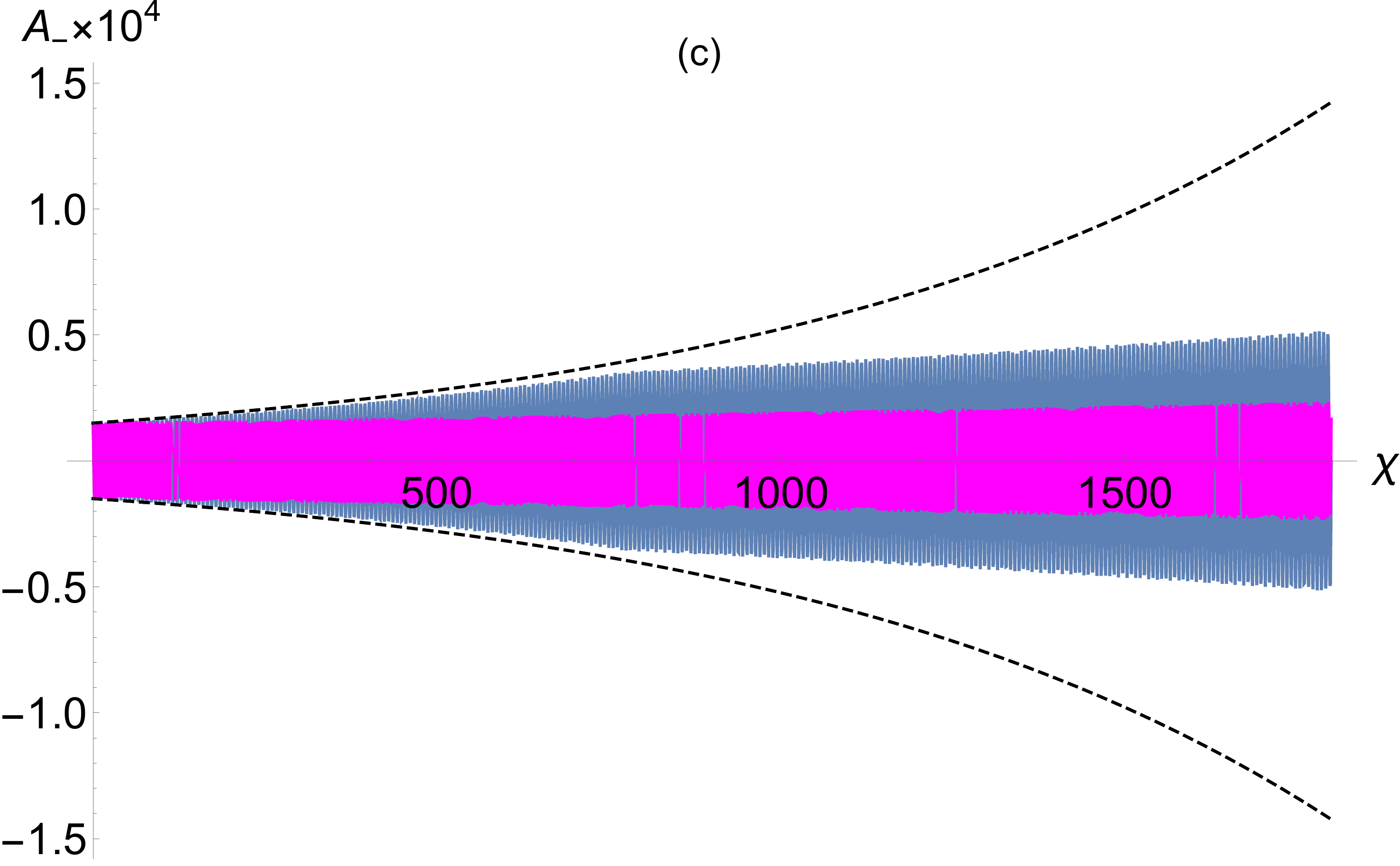}
\caption{Numerical solutions for electromagnetic plasma wave $A_-$ from Eq.~\eqref{ecGenAh}. In all figures, in dashed line is shown the theoretical exponential growing \eqref{eqexornt} for wave amplitude. (a) Solution in blue oscilating line  for $A_{-}(0)=1.5\times 10^{-5}$,  $h_0=5\times 10^{-3}$, and under condition \eqref{condition14}. In magenta line we plot the solution with same initial conditions but with
$(\Omega_p/\omega_G)^2=3/10$.
(b) Solution in blue oscilating line  for $A_{-}(0)= 10^{-5}$, $h_0=  10^{-2}$, and under condition \eqref{condition14}.  (c) Solutions with initial conditions $A_{-}(0)=1.5\times 10^{-5}$  and  $h_0=5\times 10^{-3}$, where we have considered $(\Omega_p/\omega_G)^2=1$ (blue line) and $(\Omega_p/\omega_G)^2=9/4$ (magenta line).    }
\label{figura1}
\end{figure}

In order to fully display the parametric amplification for the propagating electromagnetic plasma wave,
 Eq.~\eqref{ecGenAh} [or Eq.~\eqref{ecGenAhWH}] is solved numerically under several conditions.  In Fig.~\ref{figura1}(a) we plot the solutions for $A_-(\chi)$ in the unstable region \eqref{condition14}. We have considered initial conditions $A_{-}(0)=1.5\times 10^{-5}$, with a background gravitational wave amplitude $h_0=5\times 10^{-3}$. The (initial) amplitude of the electromagnetic wave
is chosen to be  smaller than the amplitude of the gravitational wave in order to maintain the condition of the  gravitational wave as a background for the dynamics of the electromagnetic field. The propagating oscillating solution for $A_-$ is displayed in blue solid line. We also show the exponential grow (dashed black line) of the electromagnetic wave amplitude predicted by Eq.~\eqref{eqexornt}. The electromagnetic plasma wave amplitude grows by approximately one order of magnitude by $\chi\approx 1800$.  In order to highlight the parametric resonant growing, 
in the same figure we display the solution for $A_{-}$ (with the same previous initial conditions) for a departure of the condition 
\eqref{condition14}, by choosing
$(\Omega_p/\omega_G)^2=3/10$. This solution (magenta line) represents mainly a sinoudoidal oscillation with no amplification.
 A similar behavior can be found for $A_+$.

Notice that the growth rate is quite small as compared to the oscillation frequency. It will become commensurately larger for larger $h_0$. To show this, in Fig.~\ref{figura1}(b), we plot the numerical solution for $A_-(\chi)$ in the unstable region \eqref{condition14}, with initial conditions $A_{-}(0)= 10^{-5}$ 
and   $h_0=  10^{-2}$.  For this case,  by $\chi\approx 1800$, the electromagnetic plasma wave amplitude grows by two orders of magnitude [anew, the dashed line represents the exponential grow \eqref{eqexornt}].

Finally, in Fig.~\ref{figura1}(c), we display the solutions for Eq.~\eqref{ecGenAh} for the two following unstable regions of Mathieu equation, $(\Omega_p/\omega_G)^2=1$ (blue line) and $(\Omega_p/\omega_G)^2=9/4$ (magenta line) \cite{bender}. We have used the initial conditions $A_{-}(0)=1.5\times 10^{-5}$  and $h_0=5\times 10^{-3}$.
Both electromagnetic plasma wave solutions do not present a resonant amplification, as it was discussed previously.

In general, the width of the parametric resonance is of order $h_0$; exponentially growing solutions are found only in the range \cite{bender}
\begin{equation}
 \left(\frac{\Omega_p}{\omega_G}\right)^2=\frac{1}{4} \pm \frac{h_0}{4},
\label{condition14b}
\end{equation}
beyond which pure oscillatory solutions pertain

\section{Discussion}

Since the basic physics is contained in the rather simple (highly investigated)  equation, it is no wonder that numerical and semi-analytical treatments give the same results. The most important task, however, is to dwell 
on precise conditions for the whole dynamic to take place.

The first one is the phase-matching condition in a dispersive realm. Practically, it is insured by demanding that both waves amplitudes are functions of the same propagation phase, $\omega t-kz$. It is this demand that leads to the ordinary differential equation coupling the gravitational wave to the electromagnetic wave.
This kind of phase-matching process between the electromagnetic wave, and its gravitational wave background, can only be possible for subluminal waves; both waves must be dispersive.

The second condition relates the two different (gravitational and electromagnetic) frequency responses of the medium; these must lie in the near neighborhood of $\omega_G=2 \Omega_p$ [see Eq.~\eqref{condition14} and \eqref{condition14b}]. When this condition is satisfied, even a low amplitude gravitational wave (that adds a periodic potential to the equation of the electromagnetic propagation ) can trigger a parametric instability. 


 For a medium with electron plasma, one can estimate from this condition that  $\omega_G=113,5 \sqrt{n/f}$, where $n$ is measured in m$^{-3}$. We can use this relation to examine what class of  media will  support this parametric resonance.For an electron density $n\sim 10^{30}$m$^{-3}$, and relativistic  temperatures $T\sim 10^{10}$K ($f\sim 7$), for instance, $\omega_G\sim 43$ PHz. This is a very high-frequency for electromagnetic plasma waves, on the range of ionizing radiation. Therefore, in order for a gravitational wave to trigger the parametric resonance, the frequency $\omega$ of the gravitational wave must be even larger.  For a very dilutes plasma with $n\sim 1$m$^{-3}$ with $f\sim 1$, on the contrary, it is obtained that  $\omega_G\sim 113.5$ Hz. This is well within the range of future tecnological capabilities of gravitational waves detectors- frequencies of the order $\omega\sim 300-1000$Hz \cite{ligo}.

On the other hand, under the same conditions, it is very unlikely that the reverse process takes place, i.e., a parametric amplification of gravitational waves in an electromagnetic background field. The contribution of  the electromagnetic wave to the energy momentum tensor is likely to be insignificant to affect the nature of the gravitational wave.

The transfer of energy from the sub-luminal gravitational to the sub-luminal electromagnetic wave may be one of the more significant contributors to the presence of electromagnetic energy in the universe. Though this model calculation was done for the low amplitude gravitational waves, such an exchange  is likely to happen even when the gravitational wave is very strong like in some cosmic cataclysmic events. In fact, for low amplitude waves, one can excite only the first parametric resonance. However,  higher unstable bands do
become accessible for large amplitude gravitational drive. In such a case, parametric resonance can occur for much larger ranges of $\omega_G$ and $\omega_p$. One should expect this energy transfer process between the two waves traveling in the same medium to be ubiquitous. 

Since this paper has presented a clear initial demonstration of what may turn out to be a very efficient source of electromagnetic energy, we plan to investigate the parametric resonance between sub-luminal electromagnetic  and gravitational waves in more depth and detail.


\begin{thebibliography}{}
\bibitem{footnote} In Ref.\cite {mahajan1}, one can find the electromagnetic wave interacting with the Dirac electron. The differences from the Klein-Gordon equation lie in the spin effects and are not crucial for very high energy particles.
\end{thebibliography}
\end{document}